\documentclass[12pt]{article}

\usepackage{amstext,amsmath,amssymb,amsfonts,bbm,slashed}
\usepackage[latin1]{inputenc}
\usepackage{epsfig}
\usepackage{hyperref}
\usepackage{amsthm}
\usepackage{subfigure}
\usepackage{color}
\usepackage{multirow}
\usepackage{caption}
\usepackage{graphicx}

\newcommand{\Tr}{\mathrm{Tr}}
\newcommand{\diag}{\mathrm{diag}}

\usepackage{multicol}

\usepackage[left=1.8cm,right=1.8cm,top=2cm,bottom=2.4cm]{geometry}

\newcommand{\bea}{\begin{eqnarray}}
\newcommand{\eea}{\end{eqnarray}}
\newcommand{\beq}{\begin{equation}}
\newcommand{\eeq}{\end{equation}}

\newcommand{\br}{\begin{array}}	
\newcommand{\er}{\end{array}}

\begin{document}



\vspace{20pt}

\begin{center}

\hspace{-0.5cm}
{\Large \bf
 Functional renormalization 
group for the $U(1)$-$T_5^6$ tensorial group field theory  with closure constraint 
}
\\
\vspace{0.5cm}
{\large Vincent Lahoche$^a$\footnote{vincent.lahoche@th.u-psud.fr} and Dine Ousmane Samary$^{b,c}$\footnote{dine.ousmane.samary@aei.mpg.de}}
\vspace{0.5cm}

a)\, Laboratoire de Physique Th\'eorique, CNRS-UMR 8627, Universit\'e Paris-Sud 11, 91405 Orsay Cedex, France

b)\, Max Planck Institute for Gravitational Physics, Albert Einstein Institute, Am M\"uhlenberg 1, 14476, Potsdam, Germany

c)\,  Facult\'e des Sciences et Techniques/ ICMPA-UNESCO Chair, Universit\'e d'Abomey-
Calavi, 072 BP 50, Benin

\begin{abstract}
This  paper is focused on the functional renormalization group applied to the $T_5^6$ tensor model on the Abelian  group $U(1)$ with closure constraint.  For the first time, we derive the flow equations for the
couplings and mass parameters in a suitable truncation around the marginal interactions with respect to
the perturbative power counting. For the second time, we study the behavior around the Gaussian fixed
point, and show that the theory is nonasymptotically free. Finally, we discuss the UV completion of the
theory. We show the existence of several nontrivial fixed points, study the behavior of the renormalization
group flow around them, and point out evidence in favor of an asymptotically safe theory.
\end{abstract}

\vspace{5pt}

\vspace{10pt}

\end{center}

\tableofcontents

\section{Introduction}
Tensor models (TMs) generalize matrix models and are considered as a convenient formalism for studying random
geometries \cite{Gurau:2011xp}-\cite{Gurau:2010ba}. TMs also offer an alternative to other approaches dealing with quantum gravity (QG) which is based on new mathematical/statistical tools, for example,  the $1/N$ expansion recently discovered \cite{Gurau:2011xq}-\cite{Gurau:2010ba}. On the other hand, group field theory (GFT)
is a quantum field theory over group manifolds and is considered as a second quantization of
loop quantum gravity \cite{Oriti:2014yla}-\cite{Rovelli:2010bf}. Both TMs and GFT belong to the so-called background-independent scenario for QG. They aim at describing a rudimentary phase of the geometry of spacetime, namely, when this geometry is hypothetically still in a discrete form, or at least not yet
continuous. It is also named "pre-geometric" phase of our spacetime. Recently TMs and GFT have
been combined to provide a new class of field theories so called tensorial group field theory (TGFT).
TGFTs improve the GFTs in order to allow for renormalization \cite{Carrozza:2012uv}-\cite{Geloun:2011cy}. Moreover, it has been
shown that several TGFTs models are asymptotically free in the UV, in other words, near the Gaussian fixed point
\cite{BenGeloun:2012pu}-\cite{Geloun:2016xep}.

The renormalization group (RG) method formulated first by Wilson \cite{Wilson:1971bg}-\cite{Wilson:1971dh} is a nonperturbative  method which allows to interpolate smoothly between the UV laws and the  IR phenomena in  physical systems. The RG has a vast range of applications. A particularization of the RG, the  functional renormalization group (FRG)  is a realization of the RG concept in the framework of quantum and/or statistical field theory and  is one of the best candidates for studying   quantum fluctuations \cite{Wetterich:1989xg}. An important property of this method is that the FRG could be used in regimes  where perturbative calculations are invalid,
for instance, at the vicinity of nontrivial fixed points in the infrared regime.

Recently much interest were focused on the FRG equation of various Matrix and  TGFT models \cite{Geloun:2016qyb}-\cite{Eichhorn:2013isa}.   The differential  equations of the  flow  were derived using Wetterich's equation \cite{Wetterich:1989xg}.  The  fixed points were given  and further evidence of asymptotically safety and asymptotically freedom were derived around these fixed points in the UV. 

The TGFT of the form $T^6_5$  on the $U(1)$ group with closure constraint is proved to be renormalizable   \cite{Samary:2012bw}. The proof of this claim  is performed using multi-scale analysis. The closure constraint also called gauge invariance condition can help to define the emergence of the metric on spacetime after phase transition and therefore  makes this type of model relevant for the understanding the quantum theory of  gravitation. This kind of model with closure constraint namely the six dimensional TGFT with quartic interactions is studied recently in \cite{Geloun:2016qyb} and \cite{Benedetti:2015yaa}.  The perturbative computation of the $\beta$-functions of  the $T^6_5$ model is given in \cite{Samary:2013xla}, in which, we have  showed that this model is asymptotically free in the UV. This result seems to be non consistant in the point of view of the FRG analysis. 
This paper aims at  giving the FRG analysis to a    renormalizable  tensor model $T^6_5$ with closure constraint and for improving the conclusion given in \cite{Samary:2013xla}. In a truncation containing all relevant and marginal interactions, we find  nontrivial fixed points. The   FRG flows  for coupling constants and  for the mass parameter are solved 
 numerically.

The paper is organized as follows. In section \eqref{section2} we first  present the model which is analysed  in this paper, namely the $T_5^6$  model with closure constraints. In section \eqref{sec3} we give the flow equations of the coupling constants and mass parameter  by using the dimensional renormalization parameters. In section \eqref{sec4} we give the nontrivial fixed points and  provide the numerical solution of the flow equations. In section \eqref{sec5} the validity of the choice of the truncation of the effective action is discussed. The behavior of our model in the vicinity of these fixed  points is also given. The conclusion and discussion are made in section \eqref{sec6}.

\section{The $T_5^6$ TGFT model}\label{section2}
 This section is devoted to a short review of  the particular TGFT  that we present in this paper. First, we define  and give some properties  of our model. Next, we discuss the canonical dimension that allows to make sense of the exponentiation of the functional action in the partition function.
\subsection{The model}
We consider  fields $\psi$ and $\bar \psi$ acting over the $d$ copies of the group $U(1)$, i.e.  $\psi,\,\bar\psi:\, U(1)^d\rightarrow \mathbb{C}$. 
In the Fourier representation, the fields variables $T_{\vec{p}},\, \, \bar T_{\vec p},\,\, \vec p\in \mathbb{Z}^d$  are maps $\bar{T},T:\mathbb{Z}^d \to \mathbb{C}$, such that,
\bea
\psi(\vec{\mathbf{g}})=\sum_{\vec p}T_{\vec p}\,e^{i\vec \theta \cdot \vec p},\quad \vec{\mathbf{g}}=(\mathbf{g}_1,\cdots, \mathbf{g}_d), \,\, \vec p=(p_1,\cdots,p_d),\,\vec \theta=(\theta_1,\cdots,\theta_d),\,\theta_j\in[0,2\pi).
\eea
The parameter $\theta_k$ with $\mathbf{g}_k\equiv e^{i\theta_k}$ is related to  the parametrization  of the $U(1)$ group such that $U(1)\approx S^1$. 
The   theory we consider is described by its generating function or the vacuum-vacuum transition amplitude:
\begin{equation}\label{part1}
Z_{\Lambda}[J,\bar{J}]=e^{W_\Lambda[J,\bar J]}=\int d\mu_{C_{\Lambda}}(\bar{T},T)e^{S_{int}[\bar{T},T]+\langle\bar{J},T\rangle+\langle\bar{T},J\rangle}
\end{equation}
where the notation $\langle.,.\rangle$ means: $\langle\bar{J},T\rangle=\sum_{\vec{p}\in \mathbb{Z}^d} \bar{J}_{\vec{p}}T_{\vec{p}}$, $d\mu_{C_{\Lambda}}$ is the Gaussian measure with the covariance $C_{\Lambda}$ such that:
\bea
\int \,d\mu_{C_{\Lambda}} T_{\vec{p}} \bar{T}_{\vec{p}\,'}=\frac{e^{-(\vec{p}^2+m^2)/\Lambda^2}}{\vec{p}^2+m^2}\delta\bigg(\sum_{i=1}^d p_i\bigg)\delta_{\vec{p}\vec{p}^{\prime}}=C_{\Lambda}(\vec{p},\vec{p}\,'),
\eea  
 and the delta $\delta(\sum_{i=1}^d p_i)$ implements the closure constraint, see \cite{Carrozza:2012uv, Lahoche:2015ola}.  
 $\Lambda$ is the UV cutoff which will impose that the modulus of momentum vectors remains less than $\Lambda$, namely, $|\vec p|\leq \Lambda$. We keep in mind that we will  eventually take the limit $\Lambda\rightarrow \infty$. We define a model by its action at a high (UV) energy scale. The classical action $S_{int}$ is defined as a sum of tensorial invariances \cite{Gurau:2011xp}, \cite{Rivasseau:2016rgt}:
\begin{equation}\label{int}
S_{int}[\bar{T},T]=\sum_{b\in\mathcal B}\lambda_{b}\Tr_b[\bar{T},T].
\end{equation}
A tensor invariant is a polynomial in the tensor $T$ and its conjugate $\bar T$ which is invariant under the action   of the  tensor product of
$d$ independent copies of the unitary group $U(N)$. The sum is taken over a finite set $\mathcal B$ of such invariants $d$-bubbles \cite{Gurau:2011xp} associated with the couplings $\lambda_b$.

The interaction \eqref{int} of a tensor  field theory in dimension $d=5$  \cite{Samary:2012bw} is
\bea\label{int1}
\nonumber S_{int}[\bar{T},T]=& \frac{\lambda_1}{2} \sum_{\ell = 1}^5  \sum_{\{\vec{p}_i\}\,i=1,...,4} \mathcal{W}^{(\ell)}_{\vec{p}_1,\vec{p}_2,\vec{p}_3,\vec{p}_4}T_{\vec{p}_1}\bar{T}_{\vec{p}_2}T_{\vec{p}_3}\bar{T}_{\vec{p}_4}\\
&+ \frac{\lambda_2}{3} \sum_{\ell = 1}^5 \sum_{\{\vec{p}_i\}\,i=1,...,6} \mathcal{X}^{(\ell)}_{\vec{p}_1,\vec{p}_2,\vec{p}_3,\vec{p}_4,\vec{p}_5,\vec{p}_6} T_{\vec{p}_1}\bar{T}_{\vec{p}_2}T_{\vec{p}_3}\bar{T}_{\vec{p}_4}T_{\vec{p}_5}\bar{T}_{\vec{p}_6} \cr
&+ \lambda_3 \sum_{\ell_{i}=1,i=1,2,3}^5  \sum_{\{\vec{p}_i\}\,i=1,...,6} \mathcal{Y}^{(\ell_1,\ell_2,\ell_3)}_{\vec{p}_1,\vec{p}_2,\vec{p}_3,\vec{p}_4,\vec{p}_5,\vec{p}_6}T_{\vec{p}_1}\bar{T}_{\vec{p}_2}T_{\vec{p}_3}\bar{T}_{\vec{p}_4}T_{\vec{p}_5}\bar{T}_{\vec{p}_6},
\eea
where the symbols $\mathcal{W}^{(\ell)}$, $\mathcal{X}^{(\ell)}$ and $\mathcal{Y}^{(\ell)}$ are products of delta functions associated to tensor invariant interactions, and $\lambda_i (\Lambda)$ are coupling constants. For instance:
\begin{equation}
\mathcal{W}^{(\ell)}_{\vec{p}_1,\vec{p}_2,\vec{p}_3,\vec{p}_4} = \delta_{p_{1\ell}p_{4\ell}}\delta_{p_{2\ell}p_{3\ell}}\prod_{j\neq \ell}\delta_{p_{1j}p_{2j}}\delta_{p_{3j}p_{4j}}.
\end{equation}
Such a kernel is called {\it\, bubble}  \cite{Gurau:2011xp}, and can be pictured graphically as a $6$-colored bipartite regular graph, with black and white vertices corresponding respectively to the fields $T$ and $\bar{T}$, and each line corresponding to a Kronecker delta. As an example, the $4$-valent bubble associated to the kernel $\mathcal{W}^{(1)}$ is depicted on Figure \eqref{fignew} below, and in the same way, all the interaction bubbles involved in the action $S_{int}$ are defined as:
\begin{figure}
\bea\label{bubble}
\mathcal{W}^{(\ell)}(\mathbf{g}_1,\mathbf{g}_2,\mathbf{g}_3,\mathbf{g}_4) &=& \vcenter{\hbox{\includegraphics[scale=0.8]{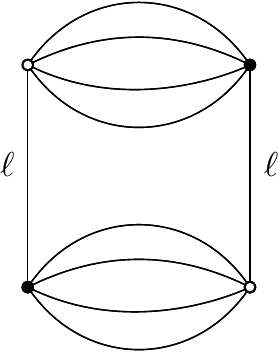}}}\\
\mathcal{X}^{(\ell)}(\mathbf{g}_1, \mathbf{g}_2, \mathbf{g}_3, \mathbf{g}_4, \mathbf{g}_5, \mathbf{g}_6) &=& \vcenter{\hbox{\includegraphics[scale=0.7]{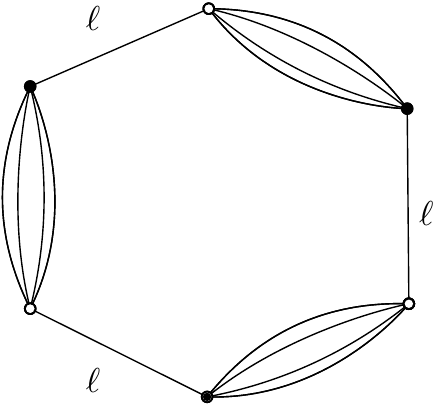}}}\\
\mathcal{Y}^{(\ell_1,\ell_2,\ell_3)}(\mathbf{g}_1, \mathbf{g}_2, \mathbf{g}_3, \mathbf{g}_4, \mathbf{g}_5, \mathbf{g}_6) &=& \vcenter{\hbox{\includegraphics[scale=0.7]{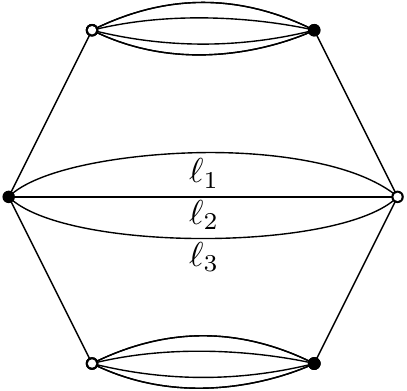}}}
\eea
\end{figure}
\begin{figure}
\begin{center}
\includegraphics[scale=0.6]{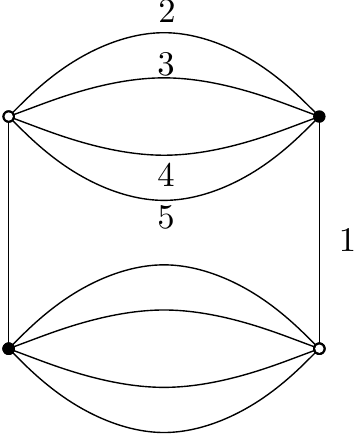}
\end{center}
\caption{An example of bubble of order 4}\label{fignew}
\end{figure}
where the index $\ell$ takes values  from $1$ to $5$, and refers to the single color characterizing each bubbles.
\subsection{Canonical dimension}
\label{sectiondim}Another important definition  for our purpose concern the notion of {\it\, canonical dimension}. We will only give the essential here, and the reader interested in the details may  consult \cite{Carrozza:2014rba}. In our model, the divergence degree for an arbitrary Feynman graph $\mathcal{G}$ is given by \cite{Carrozza:2012uv, Samary:2012bw}
\begin{equation}
\omega(\mathcal{G})=-2L(\mathcal{G})+(F(\mathcal{G})-R(\mathcal{G}))
\end{equation}
where $L$ is the number of propagators, $F$ the number of faces.  Let us pick an arbitrary orientation for all of the edges $e$ and  for all of the faces $f$. Then  $R$ is the rank of the incidence matrix $\epsilon_{fe}$: 
\bea
\epsilon_{fe}&=\begin{cases}
 \phantom{-}1& \mbox{ if $e\in f$ and their orientation match}\\
 -1 &  \mbox{ if $e\in f$  and their orientation do not match}\\
 \phantom{-}0  & \mbox{ otherwise}.
\end{cases}
\eea
Note that the rank $R$ does not depends on the chosen of the orientation. Denoting by $n_i(\mathcal{G})$ the number of  bubbles in $\mathcal{G}$ with $2i$ black and white nodes, the divergences subgraphs are said to be melonic \cite{Gurau:2011xq, Carrozza:2012uv}, if and only if they satisfy the following relation:
\begin{equation}
F(\mathcal{G})-R(\mathcal{G})=3(L(\mathcal{G})-\sum_in_i(\mathcal{G})+1)
\end{equation}
which, together with the topological relation $L(\mathcal{G})=\sum_i in_i(\mathcal{G})-N(\mathcal{G})/2$ leads to:
\begin{equation}\label{powercountingmelo}
\omega(\mathcal{G})=3-\frac{N(\mathcal{G})}{2}-2n_1(\mathcal{G})-n_2(\mathcal{G})\,,
\end{equation}
where $N(\mathcal{G})$ denote the number of external lines of $\mathcal{G}$.  For the rest  $n_1(\mathcal{G})=0$.  For $N=4$, $\omega \leq 1$, the value $1$ corresponding to melonic graphs with only 6-point interactions bubble. This conclusion indicates that perturbatively around the Gaussian Fixed Point (GFP), the coupling constant $\lambda_1$ scales as $\Lambda$ for some cut-off $\Lambda$, and we associate a canonical dimension $[\lambda_1]=1$ to this constant. In the same way, we deduce that for a generic coupling $\lambda_b$, associated to a melonic bubble with $N_b$ external lines:
\begin{equation}
[\lambda_b]=3-\frac{N_b}{2}
\end{equation}
giving explicitly:
\begin{equation}\label{dimgeneral}
[m]=1\qquad [\lambda_1]=1\qquad [\lambda_2]=[\lambda_3]=0.
\end{equation}

\section{Functional renormalization group with closure constraint}\label{sec3}
In this section we discuss  the physical consequences of the renormalization group flow by truncating the space of actions. The procedure is standard, and consists in a systematic projection of the renormalization group flow into a finite dimensional subspace of generalized couplings. The approximated trajectory $\{\Gamma_k\}$ is then described by several functions, which are solution of a finite coupled system of differential equations, the so-called {\it\, $\beta$-functions}. The difficult point of this approach is to justify the choice of the truncation. For our purpose, we use a standard dimensional argument, and neglect all the interactions up to the marginal coupling with respect to the perturbative power counting. As discussed in this section, such a truncation make sense as long as the anomalous dimension remains small, giving a consistency condition for the validity of the truncation, which can be  easily checked. However, we will see that, truncation may introduce singular artefact, as lines of fixed points, which depends on the truncation. \\

We start this section by evaluating the Wetterich equation, and find the system of $\beta$-functions studied for our model. The asymptotic behavior in the UV is also provided.
\subsection{Truncation and Regularization}
The FRG method is based on the following deformation to our original partition function given in the equation \eqref{part1} i.e.
\begin{equation}\label{family}
Z_{k}[J,\bar{J}]=\int d\mu_{C_{\Lambda}}(\bar{T},T)e^{S_{int}[\bar{T},T]-\Delta S_k[\bar{T},T]+\langle\bar{J},T\rangle+\langle\bar{T},J\rangle}
\end{equation}
where we have added to the original action a IR cut-off $\Delta S_k[\bar{T},T]$, defined as:
\begin{equation}
\Delta S_k[\bar{T},T]=\sum_{\vec{p}\in\mathbb{Z}^5}R_k(|\vec{p}|)\bar{T}_{\vec{p}}T_{\vec{p}}.
\end{equation}
The cut-off function $R_k$, depend on the real parameter $k$ playing the role of a running cut-off, and is chosen such that:\\

$\bullet$ $R_{s}(\vec{p})\geq 0$ for all $\vec{p}\in \mathbb{Z}^d$ and $s\in(-\infty,+\infty)$.\\

$\bullet$ $\lim_{s\to-\infty} R_s(\vec{p}) =  0$, implying:
$
\mathcal{Z}_{s=-\infty}[\bar{J},J]=\mathcal{Z}[\bar{J},J].
$
This condition ensures that the original model is in the family \eqref{family}. Physically, it means that the original model is recovered when all the fluctuations are integrated out.  \\

$\bullet$  $\lim_{s\to\ln\Lambda} R_s(\vec{p}) =  +\infty$, ensuring that all the fluctuations are frozen when $e^s=\Lambda$.
As a consequence, the bare action will be represented by the initial condition for the flow at $s=\ln\Lambda$.\\

$\bullet$ For $-\infty<s<\ln \Lambda$, the cutoff $R_{s}$ is chosen so that 
$
R_{s}(|p|>e^s)\ll 1,
$
a condition ensuring that the UV modes $|p|> e^s$ are almost unaffected by the additional cutoff term, while $R_{s}(|p|<e^s)\sim 1$, or  $R_{s}(|p|<e^s)\gg 1$, will guarantee that the IR modes $|p|< e^s$ are decoupled.\\

$\bullet$  $\frac{d}{d|\vec{p}|} R_s (\vec{p}) \leq 0$, for all $\vec{p}\in \mathbb{Z}^d$ and $s\in(-\infty,+\infty)$, which means that high modes should not be suppressed more than low modes.\\

The equation describing the flow of the couplings, the so called Wetterich equation has been established in \cite{Wetterich:1989xg} in the case of a theory with closure constraint:
For a given cut-off $R_k$, the effective average action satisfies the following first order partial differential equation:
\begin{equation}\label{Wettericheq}
\partial_k\Gamma_k=\sum_{\vec{p}\in\mathbb{Z}^5}\partial_kR_k(|\vec{p}|)\cdot \big[\Gamma^{(2)}_k+R_k\big]^{-1}(\vec{p},\vec{p})\delta\bigg(\sum_{i=1}^5 p_i\bigg).
\end{equation}
where $\Gamma_k$, the effective average action and is defined as the Legendre transform of the free energy $W_k:=\ln[Z_k]$ as :
\begin{equation}
\Gamma_k[\bar{T},T]+\sum_{\vec{p}\in\mathbb{Z}^5}R_k(|\vec{p}|)\bar{T}_{\vec{p}}T_{\vec{p}}:=\langle\bar{J},T\rangle+\langle\bar{T},J\rangle-W_k[J,\bar{J}]
\end{equation}
and 
\begin{equation}
\Gamma^{(2)}_k(\vec{p},\vec{p}^{\prime}):=\frac{\partial^2\Gamma_k}{\partial T_{\vec{p}}\partial \bar{T}_{\vec{p}^{\prime}}}
\end{equation}
where $T$ denote the mean field $T_{\vec{p}}:=\frac{\partial W_k}{\partial \bar{J}_{\vec{p}}}$, and is a gauge invariant field in the sens that: $T_{\vec{p}}=T_{\vec{p}}\,\delta\big(\sum_{i=1}^5 p_i\big)$.

The Wetterich flow  equation is an exact  differential
equation which must
be truncated, i.e. it must be projected to functions of few variables or even
onto some finite-dimensional sub-theory space. However as in  nonperturbative
analysis \cite{Wetterich:1989xg}, and discussed in introduction of this section the question of error estimate is very important and nontrivial in
functional renormalization. One way to estimate the error in FRG is to improve
the truncation in successive steps, i.e. to enlarge the sub-theory space
by including more and more running couplings. The difference in the flows
for different truncations gives a good estimate of the error. In addition,
one can use different regulator functions $R_k$ in a given (fixed) truncation and
determine the difference of the RG flows in the infrared for the respective
regulator choices. In this section, we adopt the simplest truncation, consisting in a restriction to the essential and marginal coupling with respect to the perturbative power counting (i.e. whose canonical dimension is upper or equal to zero). As mentioned  before, such a truncation make sense as long as the anomalous dimension remains small, and a qualitative argument is the following. Let us define the anomalous dimension $\eta:=\partial_s\ln(Z)$ (see equation \eqref{ansatz} below). In the vicinity of a fixed point, $\eta$ can reach to a non-zero value $\eta_*$. As a result, the effective propagator becomes:
\begin{equation}
\frac{Z^{-1}}{\vec{p}\,^2+(m_s^2/Z)}\approx \frac{e^{-\eta_* s}}{\vec{p}\,^2+m_*^2}\quad,
\end{equation}
and then modifies the power counting \eqref{powercountingmelo}, which becomes in the melonic sector (all the star-quantities refers to the non-Gaussian fixed point that we consider):
\begin{equation}
\omega_*(\mathcal{G})=-(2+\eta_*)L(\mathcal{G})+(F(\mathcal{G})-R(\mathcal{G}))=3-\frac{N}{2}(1-\eta_*)-3\eta_*n_3-(1+2\eta_*)n_2-(2+\eta_*)n_1\,.
\end{equation}
 As a result, the canonical dimension \eqref{dimgeneral} turn to be
\begin{equation}
[t_b]_*=3-\frac{N_b}{2}(1-\eta_*)=[t_b]+\dfrac{N_b}{2}\eta_*\quad,\label{newcandim}
\end{equation}
from which one can argue that, as long as $\eta_*\ll 1$, the classification in term of essential, inessential and marginal couplings remains unchanged, and the truncation around marginal couplings with respect to the perturbative power counting make sense.   Note that for more specific explanation the study of the critical exponent will help to prove whether or not the truncation given below equation  should be improve or not.  Unlike to what happens in a standard local field theory, each line here has several strands (the theory is non-local).
The contractions in the loop of the tadpole concerns only 4 strands out of 5. The last strand circulates freely, and
corresponds to an external momentum. It is by developing on this external variable that we generate the contribution to the anomalous dimension $\eta$. Thus, the quantity $\mathcal{W}^{(\ell)}_{\vec{p}_1,\vec{p}_2,\vec{p}_3,\vec{p}_4}$ does not explicitly depends on the momentums. The  dependence  on the momentums  is due to the non-locality of the interactions.
Up to these consideration, our choice of truncation is the following:
\bea\label{ansatz}
\nonumber \Gamma_k[\bar{T},T]=& \sum_{\vec{p}\in\mathbb{Z}^5}\bigg(Z(k)\vec{p}\,^2+m^2(k)\bigg)T_{\vec{p}}\bar{T}_{\vec{p}}+\frac{\lambda_1(k)}{2} \sum_{\ell = 1}^5  \sum_{\{\vec{p}_i\}\,i=1,...,4} \mathcal{W}^{(\ell)}_{\vec{p}_1,\vec{p}_2,\vec{p}_3,\vec{p}_4}T_{\vec{p}_1}\bar{T}_{\vec{p}_2}T_{\vec{p}_3}\bar{T}_{\vec{p}_4}\\
&\nonumber+\frac{\lambda_2(k)}{3} \sum_{\ell = 1}^5 \sum_{\{\vec{p}_i\}\,i=1,...,6} \mathcal{X}^{(\ell)}_{\vec{p}_1,\vec{p}_2,\vec{p}_3,\vec{p}_4,\vec{p}_5,\vec{p}_6} T_{\vec{p}_1}\bar{T}_{\vec{p}_2}T_{\vec{p}_3}\bar{T}_{\vec{p}_4}T_{\vec{p}_5}\bar{T}_{\vec{p}_6} \\
&+ \lambda_3(k) \sum_{\ell_i = 1,i=1,2,3}^5  \sum_{\{\vec{p}_i\}\,i=1,...,6} \mathcal{Y}^{(\ell_1,\ell_2,\ell_3)}_{\vec{p}_1,\vec{p}_2,\vec{p}_3,\vec{p}_4,\vec{p}_5,\vec{p}_6}T_{\vec{p}_1}\bar{T}_{\vec{p}_2}T_{\vec{p}_3}\bar{T}_{\vec{p}_4}T_{\vec{p}_5}\bar{T}_{\vec{p}_6}.
\eea
Also note that we have adopted an additional restriction concerning the degree of the differential operator for the kinetic term, which can be viewed as the first term in the {\it\, derivative expansion}. One more time, a consistency check must be to introduce the next contribution, and evaluate its relative contribution. We will not considered this question in this paper.\\

\noindent
We move on to the extraction of the truncated flow equations for $m^2$, $Z$ and $\lambda_i$ from the full Wetterich equation \eqref{Wettericheq}. We write the second derivative of $\Gamma_k$ as:
\begin{equation}
\nonumber\Gamma_k^{(2)}[\bar{T},T](\vec{p},\vec{p}^{\prime}) =\left( -Z(k)\vec p\,^2+m^2(k)\right)\delta\bigg(\sum_{i=1}^5p_i\bigg)\delta_{\vec{p}\vec{p}^{\prime}} + F_{k,(1)} [\bar{T},T]_{\vec{p},\vec{p}^{\prime}} + F_{k,(2)} [\bar{T},T]_{\vec{p},\vec{p}^{\prime}}
\end{equation}
in such a way that all the field-dependent terms of order $2n$ are in $F_{k, (n)}$. In particular, $F_{k,(1)}$ depends on $\lambda_1 (k)$, while $F_{k,(2)}$ depends on $\lambda_2 (k)$ and $\lambda_3 (k)$. 
\\

For the regulator $R_k$, we adopt the Litim's cut-off \cite{Litim:2000ci}, in which we set $e^k\rightarrow k$:
\begin{equation}\label{regulator}
R_k(|\vec{p}|)=Z(k)(k^2-\vec{p}\,^2)\Theta(k^2-\vec{p}\,^2),
\end{equation}
and computing the first derivative with respect to $k$, we find:
\begin{equation}\label{derivative}
k\partial_kR_k(|\vec{p}|)=\big\{k\partial_kZ(k)(k^2-\vec{p}\,^2)+2Z(k)k^2\big\}\Theta(k^2-\vec{p}\,^2).
\end{equation}
Hence, we are now in position to extract the flow equations for each couplings, which is the subject of the next section. 
\subsection{Flow equations in the UV regime}
We will deduce the flow equation in the UV regime.  In this regime, all the sums can be replaced by integration following the arguments of \cite{Benedetti:2015yaa}, essentially because the divergences of the integral approximations are the same as the exact sums. The method consists to a formal expansion of the r.h.s of the Wetterich equation \eqref{Wettericheq} in power of couplings, and identification of the corresponding terms in the l.h.s. The r.h.s involves in general some contractions between the $F_{k(n)}$ and the effective propagator $\partial_kR_k$. And in this UV regime, only the melonic graphs contribute. \\

Expanding the r.h.s and the l.h.s of the flow equation, we obtain the following relations (in matrix notations):
\begin{equation}\label{twopoints}
k\partial_k\Gamma_{k,(1)}=-\Tr_{GI}\big[\partial_k R_k \mathcal{K}_k^{-1} F_{k,(1)}\mathcal{K}_k^{-1}\big]\,,
\end{equation}
\begin{equation}\label{fourpoints}
k\partial_k\Gamma_{k,(2)}=-\Tr_{GI}\big[\partial_k R_k \mathcal{K}_k^{-1} F_{k,(2)}\mathcal{K}_k^{-1}\big] + \Tr_{GI}\big[\partial_k R_k \mathcal{K}_k^{-1} (F_{k,(1)} \mathcal{K}_k^{-1})^2\big]\,,
\end{equation}
\bea\label{sixpoints}
k\partial_k\Gamma_{k,(3)} &= 2\Tr_{GI}\big[\partial_k R_k \mathcal{K}_k^{-1} F_{k,(1)} \mathcal{K}_k^{-1}F_{k,(2)}\mathcal{K}_k^{-1}\big] 
 - \Tr_{GI}\big[\partial_k R_k \mathcal{K}_k^{-1} (F_{k,(1)} \mathcal{K}_k^{-1})^3\big]\,.
\eea
where the subscript $GI$ means "Gauge Invariant" sums, in the sense that all the terms summing involve a product with a delta $\delta\bigg(\sum_{i=1}^5p_i\bigg)$, $\Gamma_{k(n)}$ means the term of order $n$ in the truncation of equation \eqref{ansatz}, and:
\begin{equation}
\mathcal{K}_k^{-1} :=\frac{1}{Z(k)\vec{p}\,^2+m^2(k)+R_k(|\vec{p}|)}.
\end{equation}

\subsubsection{Flow equations for $Z$ and $m^2$}

\begin{figure}
\begin{center}
\includegraphics[scale=1.2]{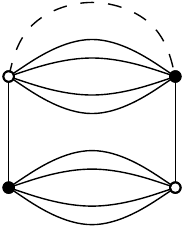} 
\caption{Typical contribution to the 2-points observable. The dashed line represents the contraction with the propagator $\partial_kR_k$.}
\label{fig2}
\end{center}
\end{figure}

Expanding the trace in the r.h.s of the equation \eqref{twopoints}, we find, using the expressions \eqref{regulator} and \eqref{derivative}
\bea
\Tr_{GI}\big[\partial_k R_k \mathcal{K}_k^{-1} F_{k,(1)}\mathcal{K}_k^{-1}\big]=\sum_{\vec{p}^{\,2}\leq k^2}\frac{k\partial_kZ(k)(k^2-\vec{p}^2)+2Z(k)k^2}{[Z(k)k^2+m^2(k)]^2}F_{k(1)}(\vec{p},\vec{p})\delta\bigg(\sum_{i=1}^5p_i\bigg)\,.
\eea
Graphically, this contribution can be pictured as in Figure \eqref{fig2}, where the dashed line represent the contraction with the propagator $\partial_kR_k$. 
The operator $F_{k(1)}$ is a sum over the intermediate colors labeling each 4-valent bubbles: $F_{k(1)}=\sum_{i=1}^5F_{k(1)}^{(i)}$, and each term contribute separately to the wave function and mass flow. As a result, we focus our attention on the computation of the trace for the case $i=1$. As explained in \cite{Benedetti:2015yaa}, we will identify the contribution to the coupling of a given bubble in the l.h.s by expanding the r.h.s around its local approximation i.e. around the value $q=0$, where $q$ denote the "external momentum" shared by the red lines in the Figure \eqref{fig2}. The two-points case is in a sense the more interesting, because for the wave function contribution, the local expansion require the first deviation to the exact local approximation, corresponding to the mass-term. This first deviation is proportional to $q^2$. Then, identifying the terms in front of each powers of $q^2$, we find: 

\begin{equation}\label{flowmass1}
k\partial_km^2(k)=-5\lambda_1(k)\frac{k\partial_kZ(k)(S_1(0)k^2-S_2(0))+2Z(k)k^2S_1(0)}{[Z(k)k^2+m^2(k)]^2}
\end{equation}
\begin{equation}\label{flowZ1}
k\partial_kZ(k)=\frac{-2\lambda_1(k)Z(k)k^2S_1^{''}}{[Z(k)k^2+m^2(k)]^2+\lambda_1(k)(S_1^{''}k^2-S_2^{''})}
\end{equation}
where the factor $5$ in front of the equation \eqref{flowmass1} takes into account the contributions for each $i$, $S_l^{''}$ denote the coefficient in front of $q^2$ in the expansion of $S_i(q)$ in power of $q$, and the sums $S_1$ and $S_2$ are :
\begin{equation}
S_{1}(q)=\sum_{\vec{p}\in\mathbb{Z}^4}\delta\bigg(\sum_{i=1}^4p_i+q\bigg)\theta(k^2-q^2-\vec{p}^{\,2}),
\end{equation}
\begin{equation}
S_{2}(q)=\sum_{\vec{p}\in\mathbb{Z}^4}\vec{p}^{\,2}\delta\bigg(\sum_{i=1}^4p_i+q\bigg)\theta(k^2-q^2-\vec{p}\,^2).
\end{equation}
Since we will be mostly interested in the large-$k$ limit, we can approximate the sums by integrals, replacing the Kronecker deltas by Dirac deltas.
The support of the integrals is in the intersection of the hyperplane of equation $q+ \sum_{l=1}^4p_l=0$ and the $4$-ball of radius $\sqrt{k^2-q^2}$. Note that the Kronecker delta of the closure constraint can be rewritten as $q+\vec{p}\cdot \vec{n}=0$, where $\vec{n}=(1,1,1,1)\in\mathbb{R}^{4}$ is the vector with all components equals to $1$. Using the rotational invariance of our integral, we can choose one of our coordinate axis to be in the direction $\vec{n}$. If we choose the axis $1$ in this direction, our constraint writes as $\delta(q+ p'_1|\vec{n}|)=\delta(q+2p'_1)=\delta(q/2+ p'_1)/2$, and we find the following integral approximation:
\bea
S_1(q)&\simeq&\frac{1}{2}\Omega_{3}\bigg[k^2-\frac{5q^2}{4}\bigg]^{\frac{3}{2}}\\
S_2(q)&\simeq&\frac{1}{2}\Big[\frac{5q^2}{4}+\frac{3}{5}\bigg(k^2-\frac{5q^2}{4}\bigg)\Big]\Omega_{3}\bigg(k^2-\frac{5q^2}{4}\bigg)^{\frac{3}{2}},
\eea
where $\Omega_{d}:=\pi^{d/2}/\Gamma(d/2+1)$ is the volume of the unit $d$-ball, with the special value $\Omega_3=4\pi/3$. Using this integral approximation, we obtain:
\begin{equation}
S_1(0)=\frac{2}{3}\pi k^3,\quad
S_2(0)=\frac{2}{5}\pi k^5
,\quad
S_1^{''}=-\frac{5}{4}\pi k
,\quad
S_2^{''}=-\frac{5}{12}\pi k^3
\end{equation}
giving:
\begin{equation}\label{flowmass2}
k\partial_km^2(k)=-\frac{4\pi}{3}\lambda_1(k)\frac{\eta(k)+5}{[Z(k)k^2+m^2(k)]^2}k^5
\end{equation}
\begin{equation}\label{flowZ2}
\eta(k)=\frac{5\pi}{2}\lambda_1(k)\frac{k^3}{[Z(k)k^2+m^2(k)]^2-\lambda_1(k)\frac{5}{6}\pi k^3}
\end{equation}
with the anomalous dimension $\eta(k)$ defined as:
\begin{equation}
\eta(k):=k\partial_k\ln(Z(k)).
\end{equation}
Note that, in this case, and for the other computations, the extraction of the local approximation in the UV limit brings up a very nice property of the melonic sector, called {\it\, traciality}. Traciality is a concept firstly introduced in a perturbative renormalization framework, ensuring that local approximation of high subgraphs make sense in the TFGT context \cite{Carrozza:2012uv}, \cite{Carrozza:2013wda}.

\subsubsection{Flow equation for $\lambda_1$}

The flow equation for $\lambda_1$ \eqref{fourpoints} involves two traces that we will compute separately. The first trace $\Tr_{GI}\big[\partial_k R_k \mathcal{K}_k^{-1} F_{k,(2)}\mathcal{K}_k^{-1}\big]$ involves two typical contributions that we have pictured in Figure \eqref{fig3} below. Note that the two diagrams pictured on this Figure have the same connectivity as the 4-valent interaction with intermediate line of color red, and we will only consider the flow equation for the coupling attached to one of the five color.\\
\begin{figure}
\begin{center}
\includegraphics[scale=1.3]{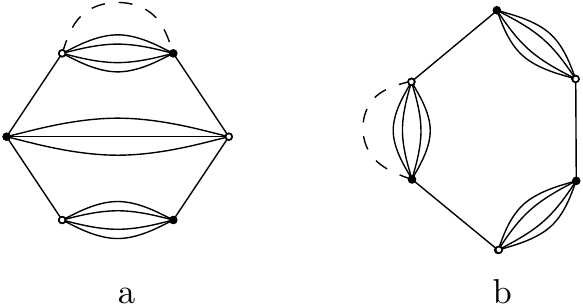}
\caption{Contributions coming from the 6-points interactions to the 4-point interaction.} \label{fig3} 
\end{center}
\end{figure}
The computation of these two contributions follows exactly the same way as the extraction of the flow equation for the mass parameter. We expand with respect to the external momentum $q$ (i.e. the momentum around the external face sharing the same line as the four internal faces) around $q=0$, the first term of the expansion giving the relevant contribution. Because the two sums involve a loop of length one, they can be expressed in terms of the two sums $S_1$ and $S_2$, and we find:
\bea\label{cont11}
\nonumber\Tr_{GI}\big[k\partial_k R_k \mathcal{K}_k^{-1} F_{k,(2)\eqref{fig3}a}\mathcal{K}_k^{-1}\big]=&4\lambda_3\frac{k\partial_kZ(k)(S_1(0)k^2-S_2(0))+2Z(k)k^2S_1(0)}{[Z(k)k^2+m^2(k)]^2}\\
&\times\sum_{\{\vec{p}_i\}\,i=1,...,4} \mathcal{W}^{(\ell)}_{\vec{p}_1,\vec{p}_2,\vec{p}_3,\vec{p}_4}T_{\vec{p}_1}\bar{T}_{\vec{p}_2}T_{\vec{p}_3}\bar{T}_{\vec{p}_4}+\mathcal{O}(q),
\eea
\bea\label{cont12}
\nonumber\Tr_{GI}\big[k\partial_k R_k \mathcal{K}_k^{-1} F_{k,(2)\eqref{fig3}b}\mathcal{K}_k^{-1}\big]=&\lambda_2\frac{k\partial_kZ(k)(S_1(0)k^2-S_2(0))+2Z(k)k^2S_1(0)}{[Z(k)k^2+m^2(k)]^2}\\
&\times\sum_{\{\vec{p}_i\}\,i=1,...,4} \mathcal{W}^{(\ell)}_{\vec{p}_1,\vec{p}_2,\vec{p}_3,\vec{p}_4}T_{\vec{p}_1}\bar{T}_{\vec{p}_2}T_{\vec{p}_3}\bar{T}_{\vec{p}_4}+\mathcal{O}(q).
\eea
The contribution of the last trace $\Tr_{GI}\big[\partial_k R_k \mathcal{K}_k^{-1} (F_{k,(1)} \mathcal{K}_k^{-1})^2\big]$ is graphically pictured in Figure \eqref{fig4}, where the dotted line means contraction with Kronecker delta. 
\begin{figure}
\begin{center}
\includegraphics[scale=1.2]{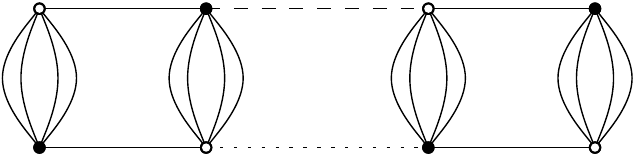} 
\caption{Contribution to the 4-point interaction involved two vertices.}\label{fig4}
\end{center}
\end{figure}
Expanding in local approximation, we find, for the extra-local contribution with $q=0$:
\begin{eqnarray}\label{cont13}
\Tr_{GI}\big[k\partial_k R_k \mathcal{K}_k^{-1} (F_{k,(1),\eqref{fig4}} \mathcal{K}_k^{-1})^2\big]&\approx&\lambda_1^2(k)\frac{k\partial_kZ(k)(S_1(0)k^2-S_2(0))+2Z(k)k^2S_1(0)}{[Z(k)k^2+m^2(k)]^3}\\
&\times&\sum_{\{\vec{p}_i\}\,i=1,...,4} \mathcal{W}^{(\ell)}_{\vec{p}_1,\vec{p}_2,\vec{p}_3,\vec{p}_4}T_{\vec{p}_1}\bar{T}_{\vec{p}_2}T_{\vec{p}_3}\bar{T}_{\vec{p}_4}.
\end{eqnarray}
Hence, summing the contributions \eqref{cont11}, \eqref{cont12} and \eqref{cont13}, and using the integral approximation for the sums, we find:
\begin{equation}\label{flowlambda12}
k\partial_k\lambda_1(k)=-(\lambda_2+4\lambda_3)\frac{4\pi}{15}\frac{\eta(k)+5}{[Z(k)k^2+m^2(k)]^2}k^5+\lambda_1^2(k)\frac{4\pi}{15}\frac{\eta(k)+5}{[Z(k)k^2+m^2(k)]^3}k^5
\end{equation}

\subsubsection{Flow equations for $\lambda_2$ and $\lambda_3$}

The only contribution to $\lambda_3$ is pictured in Figure \eqref{fig5} below, corresponding to the trace:
$$ \Tr_{GI}\big[\partial_k R_k \mathcal{K}_k^{-1} F_{k,(1)} \mathcal{K}_k^{-1}F_{k,(2)}\mathcal{K}_k^{-1}\big].$$ Indeed, it is the only contraction in the melonic sector with the same connectivity as the 6-point interaction associated with the coupling $\lambda_3$. 
\begin{figure}
\begin{center}
\includegraphics[scale=1]{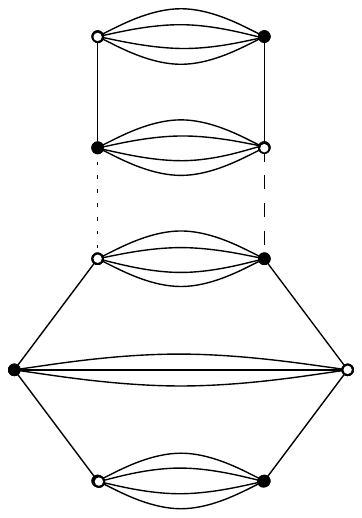} 
\caption{Contribution to the flow of $\lambda_3$}\label{fig5}
\end{center}
\end{figure}
The local approximation of the diagram can be computed exactly as for the contribution \eqref{fig3} to $\lambda_1$, and we obtain:
\bea
\Tr_{fig\,\eqref{fig5}}(q=0)&=&\lambda_1\lambda_3\frac{k\partial_kZ(k)(S_1(0)k^2-S_2(0))+2Z(k)k^2S_1(0)}{[Z(k)k^2+m^2(k)]^3}\cr
&&\times\sum_{\{\vec{p}_i\}\,i=1,...,6} \mathcal{Y}^{(\ell)}_{\vec{p}_1,\vec{p}_2,\vec{p}_3,\vec{p}_4,\vec{p}_5,\vec{p}_6}T_{\vec{p}_1}\bar{T}_{\vec{p}_2}T_{\vec{p}_3}\bar{T}_{\vec{p}_4}T_{\vec{p}_5}\bar{T}_{\vec{p}_6}
\eea
and:
\begin{equation}\label{flowlambda32}
k\partial_k\lambda_3(k)=\frac{16\pi}{15}\lambda_1\lambda_3\frac{\eta(k)+5}{[Z(k)k^2+m^2(k)]^3}k^5.
\end{equation}
In the same way, the contribution to the local approximation of the diagram \eqref{fig6}a to the flow equation for $\lambda_2$ writes as:
\bea\label{cont31}
\Tr_{fig\,\eqref{fig6}a}(q=0)&=&\lambda_1\lambda_2\frac{k\partial_kZ(k)(S_1(0)k^2-S_2(0))+2Z(k)k^2S_1(0)}{[Z(k)k^2+m^2(k)]^3}\cr
&&\times\sum_{\{\vec{p}_i\}\,i=1,...,6} \mathcal{X}^{(\ell)}_{\vec{p}_1,\vec{p}_2,\vec{p}_3,\vec{p}_4,\vec{p}_5,\vec{p}_6}T_{\vec{p}_1}\bar{T}_{\vec{p}_2}T_{\vec{p}_3}\bar{T}_{\vec{p}_4}T_{\vec{p}_5}\bar{T}_{\vec{p}_6}
\eea

\begin{figure}
\begin{center}
\includegraphics[scale=1]{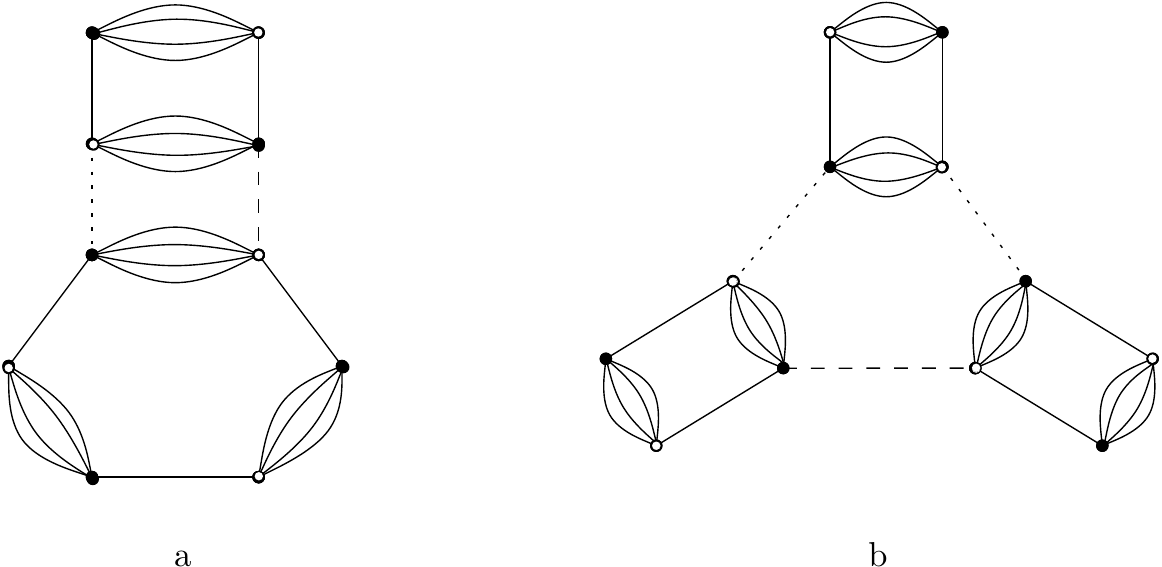} 
\caption{Contribution to the flow of $\lambda_2$}\label{fig6}
\end{center}
\end{figure}
Finally, the contribution coming from the diagram \eqref{fig6}b in Figure \eqref{fig6} involves a loop with two delta propagators, and:
\bea\label{cont32}
\Tr_{fig\,\eqref{fig6}b}(q=0)&=&\lambda_1^3(k)\frac{k\partial_kZ(k)(S_1(0)k^2-S_2(0))+2Z(k)k^2S_1(0)}{[Z(k)k^2+m^2(k)]^4}\cr
&&\times\sum_{\{\vec{p}_i\}\,i=1,...,6} \mathcal{X}^{(\ell)}_{\vec{p}_1,\vec{p}_2,\vec{p}_3,\vec{p}_4,\vec{p}_5,\vec{p}_6}T_{\vec{p}_1}\bar{T}_{\vec{p}_2}T_{\vec{p}_3}\bar{T}_{\vec{p}_4}T_{\vec{p}_5}\bar{T}_{\vec{p}_6}.
\eea
Grouping together the contributions \eqref{cont31} and \eqref{cont32}, we find:
\begin{equation}\label{flowlambda22}
k\partial\lambda_2=\frac{24\pi}{15}\lambda_1\lambda_2\frac{\eta(k)+5}{[Z(k)k^2+m^2(k)]^3}k^5-\frac{12\pi}{15}\lambda_1^3\frac{\eta(k)+5}{[Z(k)k^2+m^2(k)]^4}k^5.
\end{equation}

\subsubsection{Dimensionless renormalized parameters}
Taking into account the canonical dimension define in Section \eqref{sectiondim}, the renormalized dimensionless couplings are defined as:
\begin{equation}
m=\sqrt{Z}k\bar{m}\qquad {\lambda}_1=Z^2k\bar{\lambda}_1\qquad \lambda_2=Z^3\bar{\lambda}_2\qquad \lambda_3=Z^3\bar{\lambda}_3
\end{equation}
Using the flow equations \eqref{flowZ2}, \eqref{flowmass2}, \eqref{flowlambda12}, \eqref{flowlambda22} and \eqref{flowlambda32}, we find for the dimensionless renormalized couplings the following autonomous system:
\begin{equation}\label{flowZ3}
\eta(k)=\frac{5\pi}{2}\bar{\lambda}_1(k)\frac{1}{[1+\bar{m}^2(k)]^2-\bar{\lambda}_1(k)\frac{5}{6}\pi}
\end{equation}
\begin{equation}\label{flowmass3}
\beta_{m^2}=-(2+\eta)\bar{m}^2(k)-\frac{4\pi}{3}\bar{\lambda}_1(k)\frac{\eta(k)+5}{[1+\bar{m}^2(k)]^2}
\end{equation}
\begin{equation}\label{flowlambda13}
\beta_{\lambda_1}=-(1+2\eta)\bar{\lambda}_1(k)-(\bar{\lambda}_2+4\bar{\lambda}_3)\frac{4\pi}{15}\frac{\eta(k)+5}{[1+\bar{m}^2(k)]^2}+\bar{\lambda}_1^2(k)\frac{4\pi}{15}\frac{\eta(k)+5}{[1+\bar{m}^2(k)]^3}
\end{equation}
\begin{equation}\label{flowlambda33}
\beta_{\lambda_3}=-3\eta\bar{\lambda}_3(k)+\frac{16\pi}{15}\bar{\lambda}_1\bar{\lambda}_3\frac{\eta(k)+5}{[1+\bar{m}^2(k)]^3}.
\end{equation}
\begin{equation}\label{flowlambda23}
\beta_{\lambda_2}=-3\eta\bar{\lambda}_2(k)+\frac{24\pi}{15}\bar{\lambda}_1\bar{\lambda}_2\frac{\eta(k)+5}{[1+\bar{m}^2(k)]^3}-\bar{\lambda}_1^3\frac{12\pi}{15}\frac{\eta(k)+5}{[1+\bar{m}^2(k)]^4}\,,
\end{equation}
with the definition : $\beta_i:=k\partial_k \bar{i}$, $i\in\{m^2,\lambda_1,\lambda_2,\lambda_3\}$.
\section{Fixed points in the UV regime}\label{sec4}
At vanishing  $\beta$-functions  we obtain a fixed points. 
But these fixed points does not get any quantum correction and is called the Gaussian fixed point. In the  
neighborhood of these fixed points, the stability is determined by the linearized system of $\beta$-functions. 
All these points are studied in detail in this section.

\subsection{Vicinity of the Gaussian fixed point}

The autonomous system describing the flow of the dimensionless couplings admits a trivial fixed point for the values $\bar{\lambda}_1=\bar{\lambda}_2=\bar{\lambda}_3=\bar{m}=0$ called Gaussian Fixed Point (GFP). Expanding our equations around this points, we find the reduced autonomous system:
\bea
\begin{cases}
\phantom{-}\beta_{m^2}&\approx-2\bar{m}^2-\frac{20\pi\bar{\lambda}_1}{3}\,,\\
\phantom{-}\beta_{\lambda_1}&\approx-\bar{\lambda}_1-\frac{4\pi}{3}(\bar{\lambda}_2+4\bar{\lambda}_3)\big(1+\frac{\pi}{2}\bar{\lambda}_1+2\bar{m}^2\big)-\frac{11\pi}{3}\bar{\lambda}_1^2\,,\\
\phantom{-}\beta_{\lambda_2}&\approx\frac{\pi}{2}\bar{\lambda}_1\bar{\lambda}_2\,,\\
\phantom{-}\beta_{\lambda_3}&\approx -\frac{13\pi}{6}\bar{\lambda}_1\bar{\lambda}_3\,
\end{cases}
\label{system1}
\eea
and the anomalous dimension:
\begin{equation}\label{GFP1}
\eta(k)\approx \frac{5\pi\bar{\lambda}_1}{2}\,.
\end{equation}
These equations give the qualitative behavior of the RG trajectories around the GFP. In order to study its stability, we compute the {\it\, stability matrix} $\beta_{ij}:=\partial_i\beta_j\, i\in\{m^2,\lambda_1,\lambda_2,\lambda_3\}$, and evaluate each coefficient at the GFP. We find:
\begin{equation}\beta^{GFP}_{ij}:=\begin{pmatrix}
-2&0&0&0\\
-\frac{20\pi}{3}&-1&0&0\\
0&-\frac{4\pi}{3}&0&0\\
0&-\frac{16\pi}{3}&0&0
\end{pmatrix},
\end{equation}
with eigenvalues $(-2,-1,0,0)$ and eigenvectors $e_1^{GFP}=(\frac{9}{160\pi^2},\frac{3}{8\pi},\frac{1}{4},1); \,$ $e_2^{GFP}=(0,\frac{3}{16\pi},\frac{1}{4},1);\,$ $e_3^{GFP}=(0,0,0,1);\,$ $e_4^{GFP}=(0,0,1,0)$. One recall that the {\it\, critical exponents} are the opposite values of the eigenvalues of the $\beta_{ij}$, and that the fixed point can be classified following the sign of their critical exponents. Hence, we have two relevant directions in the UV, with critical exponents $2$ and $1$, and two marginal couplings with zero critical exponents. Moreover, note that the critical exponents are equal to the canonical dimension around the GFP. Finally, note that the previous system of equations admits other fixed points, or a {\it\, line of fixed points} in addition to the Gaussian one, for the values: $\bar{\lambda}_1=\bar{m}=0$; $\bar{\lambda}_3=-\bar{\lambda}_2/4$. This fixed point occurs as well as in the non-perturbation analysis, and we will return on this subject in the next Section.\\

\begin{figure}[h!]
\begin{minipage}[c]{.50\linewidth}
\centering
\includegraphics[scale=0.5]{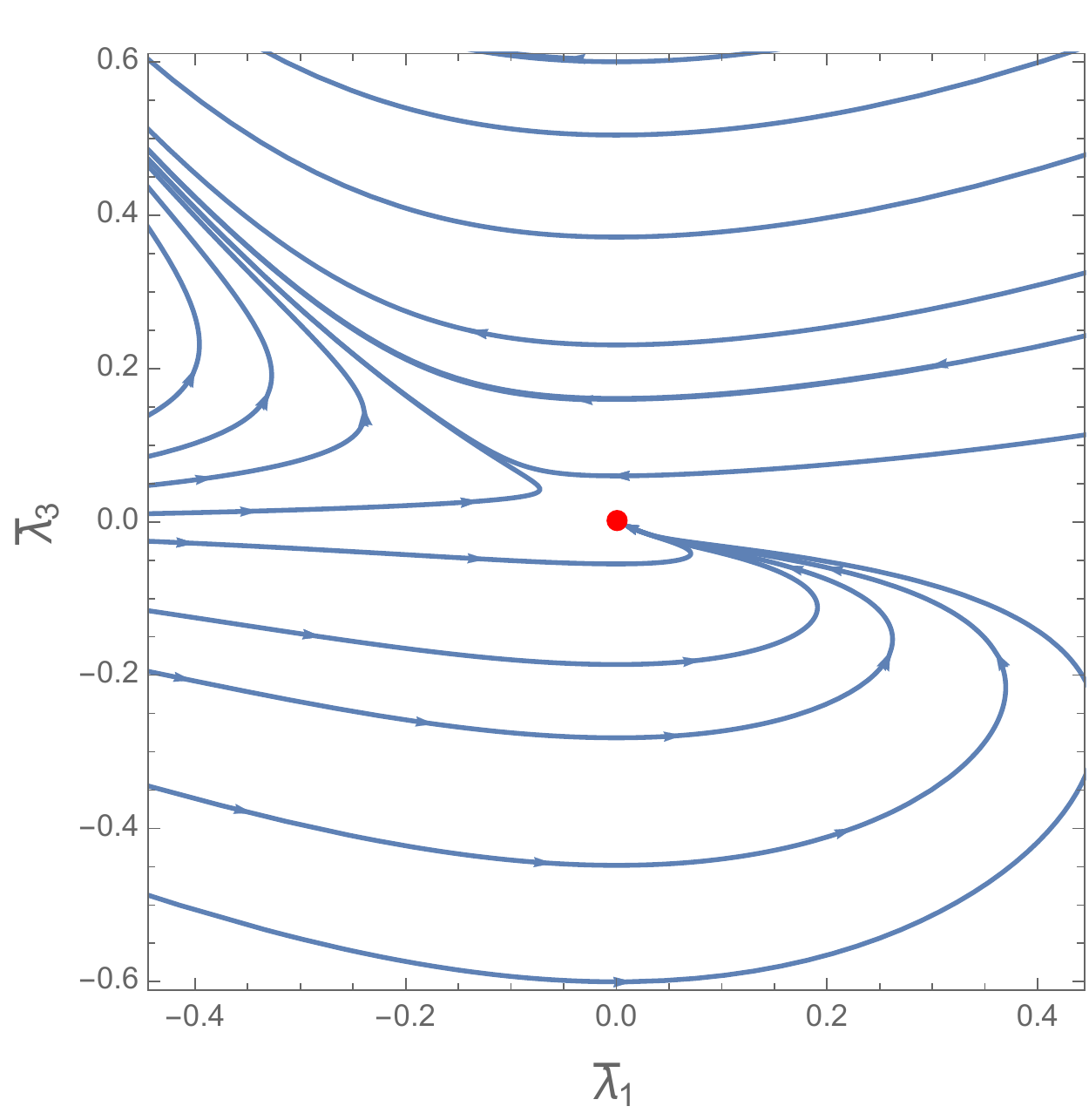}
\end{minipage}
\hfill%
\begin{minipage}[c]{.50\linewidth}
\centering
\includegraphics[scale=0.5]{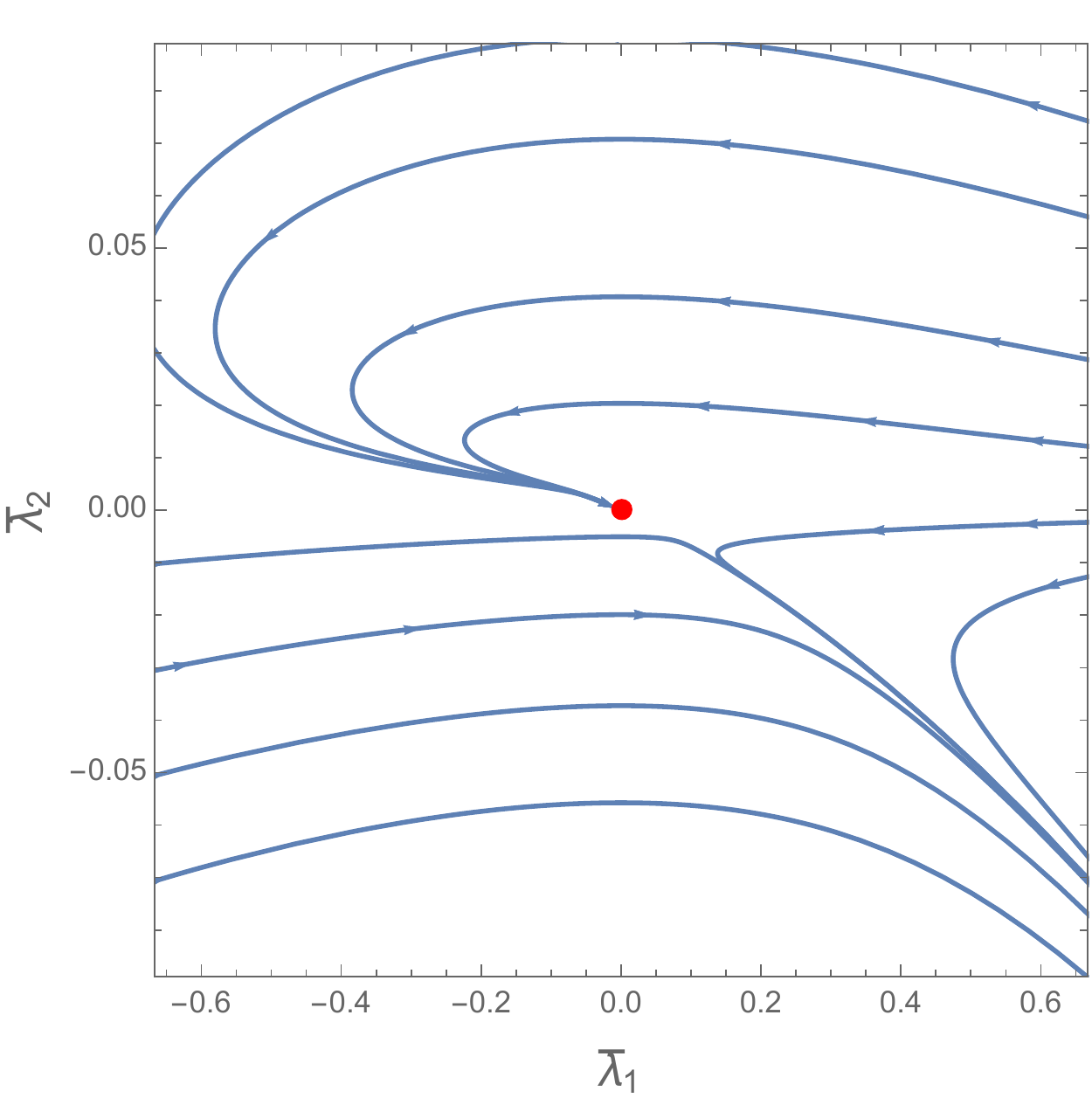}
\end{minipage}
\caption{Phase portrait in the plans $(\bar{\lambda}_1,\bar{\lambda}_3)$ for $
\bar{\lambda}_2=0$ (one the left) and in the plan $(\bar{\lambda}_1,\bar{\lambda}_2)$, for $\bar{\lambda}_3=0$.} \label{figAs}
\end{figure}

\noindent
For the moment, we are in position to discuss the qualitative flow diagram around the Gaussian fixed point. First of all, note that all the coefficients of the beta function of the system \eqref{system1} are not negative definite. This fact seems to be a special feature of this model, meaning that the weight of the anomalous dimension does not dominate the vertex contribution. This fact is a first difference with respect to the similar non-Abelian $\phi^6$ model studied in \cite{Carrozza:2014rba}. However, the analysis providing in this reference remains true, and the model is not asymptotically free. We will not repeat the complete analysis given in \cite{Carrozza:2014rba} , but a qualitative argument is the following. Exploiting the fact that the hyperplans $\bar{\lambda}_2=0$ and $\bar{\lambda}_3=0$ are invariant under the flow, we can look at only a two-dimensional reduction of the complete system \eqref{system1}. We choose $\bar {\lambda}_2=0$, and plot the numerical integration of the reduced flow equation in Figure \eqref{figAs} (on the left) below. 
In the domain, $\bar{\lambda}_3>0$, even if a given trajectory approaches of the Gaussian fixed point, $\bar{\lambda}_1$ reaches a negative value, and it is ultimately repelled for $k$ sufficiently large. The same phenomenon occurs for $\bar{\lambda_2}$ in the plan $\bar{\lambda}_2<0$ (see the Figure\eqref{figAs} one the right). The issue of the UV completion of a theory which is non-Asymptotically free is one of the difficult question that we address to the non-perturbative renormalization group machinery, and the rest of this paper is essentially devoted to this one. 

\subsection{Non-Gaussian fixed points}

Solving numerically the systems \eqref{flowmass2}-\eqref{flowlambda32}, we find some non-Gaussian fixed points, whose relevant characteristics are summarized in the Table \eqref{table1} below. In addition to  these non-Gaussian fixed points, the system admits a {\it\, line of fixed points}, $LFP$, for the  values:
\begin{equation}
LFP=\{\bar{m}^2=0,\bar{\lambda}_1=0,\bar{\lambda}_2=-4\bar{\lambda}_3\}\,,
\end{equation}
with critical exponents:
\bea
\begin{cases}
\phantom{-}\theta^{(1)}&=-2\,,\\
\phantom{-}\theta^{(2)}&=\,\,\,\,\,0\,,\\
\phantom{-}\theta^{(3)}&=-\frac{1}{2}\big(1+\sqrt{1-\frac{128}{9}\pi^2\bar{\lambda}_2}\big)\,,\\
\phantom{-}\beta_{\lambda_3}&= -\frac{1}{2}\big(1-\sqrt{1-\frac{128}{9}\pi^2\bar{\lambda}_2}\big).\,
\end{cases}
\label{system12}
\eea
\begin{table}
\begin{center}
\begin{tabular}{|l|l|l|l|l|l|l|l|l|l|}
\hline FP & $\bar{m}^2$&$\bar{\lambda}_{1}$&$\bar{\lambda}_2$ &  $\bar{\lambda}_3$ &  $\eta$&$\theta^{(1)}$&$\theta^{(2)}$&$\theta^{(3)}$&$\theta^{(4)}$\\
\hline  $FP_1$ &-0.3 &0.005& 0.0009 & -0.0002 & -6.3&-299&56.1&-11.7&5.8\\
\hline  $FP_2$ &-0.7&0.008&0.0006&-0.0002&0.76&-7.4-1.9i&-7.4+1.9i&3.34&-0.12\\
\hline  $FP_3$ &-0.9&0.0007& 3.32.$10^{-6}$& 0.& 1.3&-66.7&-42.63&-27.7&1.80\\
\hline  $FP_4$ &-0.8&0.04& -0.02 & 0. & -5.9&-144.8&-14.4&-7.5&-5.4\\
\hline  $FP_5$ &0.06&-0.006& 0.002 & 0. & -0.04&1.9&1.09&-0.04&-0.01\\
\hline  $FP_6$ &1.32&-0.5& -0.06 & 0. & -0.6&3.0&-1.23&-1.13&-0.39\\
\hline
\end{tabular}
\caption{Summary of the properties of the non-Gaussian fixed points. Again, the critical exponents $\theta^{i}$ are the opposite values of the eigenvalues of the stability matrix: $\beta_*=:\diag(-\theta^1_*,-\theta^2_*,-\theta^3_*,-\theta^4_*)$.}\label{table1}
\end{center}
\end{table}
The denominator of $\eta$, $D:=[1+\bar{m}^2(k)]^2-\bar{\lambda}_1(k)\frac{5}{6}\pi$ introduce a singularity in the flow. At the Gaussian fixed point, and in a sufficiently small domain around, $D>0$. But further away from the GFP, $D$ may cancel, creating in the $(\bar{\lambda}_1,\bar{m}^2)$-plan a singularity line. The area below this line where $D <0$ is thus disconnected from the region $D> 0$ connected to GFP. Then, we ignore for our purpose the fixed points in the disconnected region, for which $D<0$. A direct computation show that only the fixed points $FP_2$, $FP_3$, $FP_5$ and $FP_6$ are relevant for an analysis in the domain connected to the Gaussian fixed point. \\

\begin{figure}
\begin{center}
\includegraphics[scale=1.2]{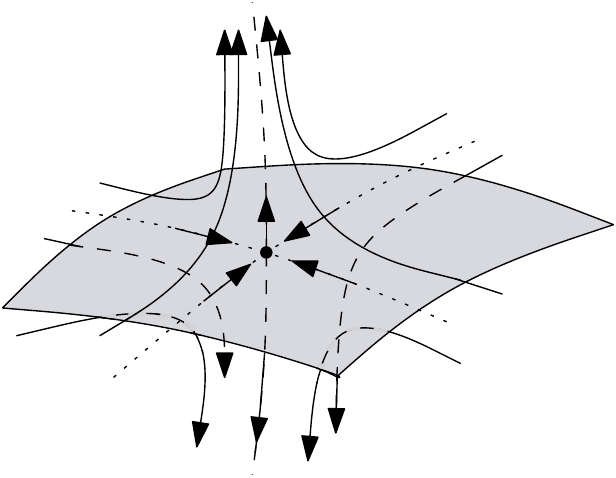} 
\caption{Qualitative behavior of the RG trajectories around an IR fixed point. The critical surface  is spanned by the relevant directions in the IR, and the arrows are oriented toward the IR direction. This illustrates the senario of  asymptotically safety. }\label{figcritical}
\end{center}
\end{figure}

\noindent
$\bullet$ The fixed points $FP_2$ and $FP_3$ are very similar. They have three irrelevant directions and one relevant direction in the UV. For each of these fixed points, the three irrelevant directions span a three dimensional manifold on which the trajectory runs toward the fixed point in the IR, while the trajectories out side are repelled of this critical surface, as pictured on Figure \eqref{figcritical}. This picture, the existence of a separatrix between two connected regions of the phase space is reminiscent of a critical behavior, with phase transition between a broken and a symmetric phase, and these separatrix are {\it\, IR-critical surfaces}. This interpretation is highlighted for the two fixed points in the zero momenta limit. Indeed, in both cases, the contributions in the effective action of the terms proportionals to $\bar{\lambda}_2$ and $\bar{\lambda}_3$ can be neglected in comparison with the contributions of the two first terms, leading in the first approximation a Ginsburg-Landau equation for $\phi^4$ scalar complex theory. Note that for $FP_2$ two critical exponents are complex, providing some oscillations of the trajectories, and implying that the fixed point is an IR-attractor in the two-dimensional manifold spanned by the eigenvectors corresponding to these two critical exponents. Moreover, the fixed point $FP_6$ appears to be an IR fixed point, with coordinates of opposite sign. \\

\noindent
$\bullet$ The fixed point $FP_5$ has two relevant and two irrelevant directions in the UV. The relevant directions in the UV span a two-dimensional manifold corresponding to a {\it\, UV-multicritical surface}. Such a surface is interesting for the UV-completion of the theory. Indeed, all the trajectories in the surface are oriented toward the fixed point in the UV, while the dimension of the surface give an interesting number of physical parameter, providing an evidence in favor of the {\it\, asymptotic safety} of the model in the UV \cite{safety}. 

\noindent
$\bullet$ Finally, we have the line of fixed point, for which we will distinguish four cases:\\

\noindent
{\it\, i}\,\, In the domain $d_1=\{\bar{\lambda}_2<0\}$ we have two relevant, one marginal and one irrelevant directions. \\

\noindent
{\it\, ii}\,\, At the point $d_2=\{\bar{\lambda}_2=0\}$, we recover the GFP, with two relevant and two marginal directions. \\

\noindent 
{\it\, iii}\,\, In the domain $d_3=\{\bar{\lambda}_2\in]0,\big(\frac{3}{8\pi}\big)^2]\,\}$ we have three relevant and one marginal directions. One more time, this section of the critical line is interesting in view of the UV-completion of the theory and provide a supplementary evidence if favor of asymptotic safety. Indeed, in each points, the relevant directions in the UV span a three dimensional UV-critical surface, in favor of the existence of a non-trivial asymptotically safe theory with three independent physical parameters. This line of fixed point has been recently discussed in \cite{Geloun:2016xep} for a similar model improved by unconnected interaction bubbles. \\

\noindent
{\it\, iv}\,\, In the domain $d_4=\{\bar{\lambda}_2>\big(\frac{3}{8\pi}\big)^2\,\}$ The situation is very reminiscent of the previous one. We have three eigenvalues with negative real part and one equal to zero. Hence, we have three relevant and one marginal directions. The only difference in comparison with the domain $d_3$ is that the eigenvalue have non-zero imaginary parts, giving some oscillations and attractor phenomena in the trajectories. \\

\noindent
Finally, we briefly discuss the values of the anomalous dimensions. With our conventions, the couplings of the relevant operator are suppressed as a power of $k$ in the UV limit $k\rightarrow \infty$. The  couplings decrease when the trajectory goes away from the UV regime. However, the power law behavior is limited to the attractive region of the fixed point, far from its scaling regime it can deviate from the power law one. And we can evaluate this deviation. For instance, in the vicinity of $FP_5$, one deduce from \eqref{newcandim} that the canonical dimension becomes:
\begin{equation}
[t_b]_{FP_5}\approx 3-1.6\frac{N_b}{2}\,,
\end{equation}
from which we deduce that all the interaction of valence up or equal to four becomes inessentials. The same phenomenon occurs in the vicinity of $FP_4$, where all the interactions up to these of valence four become inessentials. In contrary, at the fixed points $FP_2$ and $FP_4$ the anomalous dimension is positive, meaning that the power counting {\color{blue}is} improved with respect to the Gaussian one, and irrelevent operators are enhanced in the UV.

\section{Truncation with   an interaction of valence 8}\label{sec5}
\begin{figure}[htbp]
\begin{center}
 $V_{4,1}$
\includegraphics[scale=0.6]{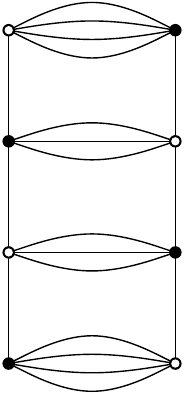}\hspace{1.cm}
 $V_{4,2}$
\includegraphics[scale=0.6]{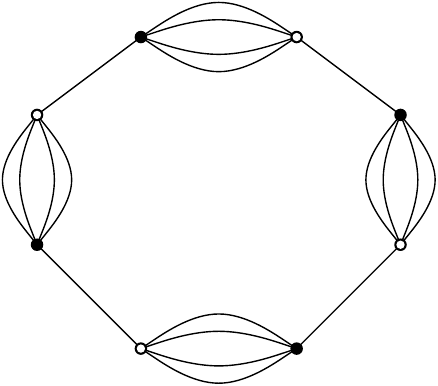}\hspace{0.5cm}
$V_{4,3}$\includegraphics[scale=0.6]{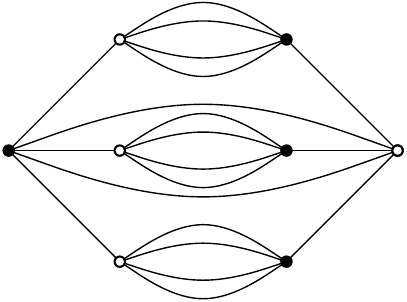}\hspace{1.cm}
$V_{4,4}$
\includegraphics[scale=0.6]{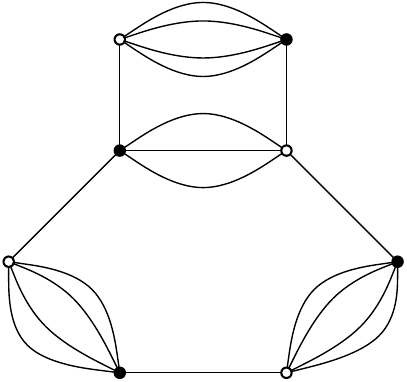}\hspace{0.5cm}
\end{center}
 \caption{Graphical representation of the vertices  of  valence 8 }
  \label{fig:Vertex4} 
\end{figure}

This section aims to identify how the adding of interaction of the valence $8$  may modify the flow equation and the fixed point of our model. This means that the effective action is now truncate to satisfy the following  form: let $e_{ji}\in [1,5]$ is the color of the bubble of valence 8, with $i=1,2,3,4$ and $j=1,2,\cdots ,5,$

\bea\label{ansatz22}
\Gamma_k[\bar{T},T]&=& \sum_{\vec{p}\in\mathbb{Z}^5}\bigg(Z(k)\vec{p}\,^2+m^2(k)\bigg)\bar{T}_{\vec{p}}T_{\vec{p}}+\frac{\lambda_1(k)}{2} \sum_{\ell = 1}^5  \sum_{\{\vec{p}_i\}\,i=1,...,4} \mathcal{W}^{(\ell)}_{\vec{p}_1,\vec{p}_2,\vec{p}_3,\vec{p}_4}T_{\vec{p}_1}\bar{T}_{\vec{p}_2}T_{\vec{p}_3}\bar{T}_{\vec{p}_4}+\cr
&+&\frac{\lambda_2(k)}{3} \sum_{\ell = 1}^5 \sum_{\{\vec{p}_i\}\,i=1,...,6} \mathcal{X}^{(\ell)}_{\vec{p}_1,\vec{p}_2,\vec{p}_3,\vec{p}_4,\vec{p}_5,\vec{p}_6} T_{\vec{p}_1}\bar{T}_{\vec{p}_2}T_{\vec{p}_3}\bar{T}_{\vec{p}_4}T_{\vec{p}_5}\bar{T}_{\vec{p}_6} \cr
&+& \lambda_3(k) \sum_{\ell_i = 1,i=1,2,3}^5  \sum_{\{\vec{p}_i\}\,i=1,...,6} \mathcal{Y}^{(\ell_1,\ell_2,\ell_3)}_{\vec{p}_1,\vec{p}_2,\vec{p}_3,\vec{p}_4,\vec{p}_5,\vec{p}_6}T_{\vec{p}_1}\bar{T}_{\vec{p}_2}T_{\vec{p}_3}\bar{T}_{\vec{p}_4}T_{\vec{p}_5}\bar{T}_{\vec{p}_6}
\cr
&+&\sum_{i=1}^4  \lambda_{4,i}(k) \sum_{e_{ji}=1/,e_{ji}\neq e_{ki} \forall j\neq k}^5  \sum_{\{\vec{p}_l\}\,l=1,...,8} \mathcal{Z}^{i,(e_{ji})}_{\vec{p}_1,\vec{p}_2,\vec{p}_3,\vec{p}_4,\vec{p}_5,\vec{p}_6,\vec{p}_7,\vec{p}_8}T_{\vec{p}_1}\bar{T}_{\vec{p}_2}T_{\vec{p}_3}\bar{T}_{\vec{p}_4}T_{\vec{p}_5}\bar{T}_{\vec{p}_6} T_{\vec{p}_7}\bar{T}_{\vec{p}_8},
\eea
where we assume that the last term of the action \eqref{ansatz22} takes into account all contributions of melonic interactions of the form $T^8$, and the coupling constants $\lambda_{4,i},\,\,i=1,2,3,4$ are related to the vertex $V_{4,i}$ see figures \eqref{fig:Vertex4}. The set $\{e_{\cdot\, i}\}_{j}$ takes into account all the colors associated to the vertices $V_{4,i}$.
We get for $\Gamma_{k,(3)}$ and  $\Gamma_{k,(4)}$ the flow equations:
\bea\label{sixpoints122}
k\partial_k\Gamma_{k,(3)} &=& 2\Tr_{GI}\big[\partial_k R_k \mathcal{K}_k^{-1} F_{k,(1)} \mathcal{K}_k^{-1}F_{k,(2)}\mathcal{K}_k^{-1}\big]  - \Tr_{GI}\big[\partial_k R_k \mathcal{K}_k^{-1} (F_{k,(1)} \mathcal{K}_k^{-1})^3\big]\cr
&&- \Tr_{GI}\big[\partial_k R_k \mathcal{K}_k^{-1} F_{k,(3)} \mathcal{K}_k^{-1}\big]\,,
\eea
and
\bea\label{4points12}
k\partial_k\Gamma_{k,(4)}&=&- \Tr_{GI}\big[\partial_k R_k \mathcal{K}_k^{-1} F_{k,(4)} \mathcal{K}_k^{-1}\big]+ \Tr_{GI}\big[\partial_k R_k \mathcal{K}_k^{-1} (F_{k,(2)} \mathcal{K}_k^{-1})^2\big]\cr
&&+ \Tr_{GI}\big[\partial_k R_k \mathcal{K}_k^{-1} (F_{k,(1)} \mathcal{K}_k^{-1})^4\big]\cr
&&-3\Tr_{GI}\big[\partial_k R_k \mathcal{K}_k^{-1} F_{k,(2)} \mathcal{K}_k^{-1}(F_{k,(1)}\mathcal{K}_k^{-1})^2\big],
\eea
Taking into account the dimensionless renormalized parameter, and grouping all melonic contributions,
the flow equations of the coupling constants $\bar \lambda_2$, $\bar \lambda_3$ and $\bar\lambda_{4,i}$, are:
\bea
\begin{cases}
 \phantom{-}
\beta_{m^2}&=-(2+\eta)\bar{m}^2-\frac{4\pi}{3}\bar{\lambda}_1\frac{\eta+5}{[1+\bar{m}^2]^2}\,,\\
\phantom{-}\beta_{\lambda_1}&=-(1+2\eta)\bar{\lambda}_1-(\bar{\lambda}_2+4\bar{\lambda}_3)\frac{4\pi}{15}\frac{\eta+5}{[1+\bar{m}^2]^2}+\bar{\lambda}_1^2\frac{4\pi}{15}\frac{\eta+5}{[1+\bar{m}^2]^3}\,,\\
\phantom{-}\beta_{\lambda_2}&=-3\eta\bar{\lambda}_2+\frac{24\pi}{15}\bar{\lambda}_1\bar{\lambda}_2\frac{\eta+5}{[1+\bar{m}^2]^3}-
\bar{\lambda}_1^3\frac{12\pi}{15}\frac{\eta+5}{[1+\bar{m}^2]^4}-\frac{48\pi}{15}\bar\lambda_{4,2}\frac{\eta+5}{[1+\bar m^2]^2}
-\frac{12\pi}{15}\bar\lambda_{4,4}\frac{\eta+5}{[1+\bar m^2]^2}\\
\phantom{-}\beta_{\lambda_3}&= -3\eta\bar{\lambda}_3+\frac{16\pi}{15}\bar{\lambda}_1\bar{\lambda}_3\frac{\eta+5}{[1+\bar{m}^2]^3}-\frac{8\pi}{15}\bar\lambda_{4,1}\frac{\eta+5}{[1+\bar m^2]^2}-\frac{36\pi}{15}\bar\lambda_{4,3}\frac{\eta+5}{[1+\bar m^2]^2}-\frac{8\pi}{15}\bar\lambda_{4,4}\frac{\eta+5}{[1+\bar m^2]^2}\,,\\
\phantom{-}\beta_{\lambda_{4,1}}&=(1-4\eta)\bar{\lambda}_{4,1}+\frac{16\pi}{15}\bar\lambda_{4,1}\bar\lambda_1\frac{\eta+5}{[1+\bar m^2]^2}+\frac{4\pi}{15}\bar\lambda_3^2\frac{\eta+5}{[1+\bar m^2]^3}-\frac{24\pi}{15}\bar\lambda_1^2\bar{\lambda}_3\frac{\eta+5}{[1+\bar m^2]^4}+\frac{4\pi}{15}\bar\lambda_1^4\frac{\eta+5}{[1+\bar m^2]^5}\,,\\
\phantom{-}\beta_{\lambda_{4,2}}&=(1-4\eta)\bar{\lambda}_{4,2}+\frac{32\pi}{15}\bar\lambda_{4,2}\bar\lambda_1\frac{\eta+5}{[1+\bar m^2]^2}
+\frac{4\pi}{15}\bar\lambda_2^2\frac{\eta+5}{[1+\bar m^2]^3}-\frac{12\pi}{15}\bar\lambda_1^2\bar\lambda_2\frac{\eta+5}{[1+\bar m^2]^4}\,,\\
\phantom{-}\beta_{\lambda_{4,3}}&=(1-4\eta)\bar{\lambda}_{4,3}+\frac{24\pi}{15}\bar\lambda_{4,3}\bar\lambda_{1}\frac{\eta+5}{[1+\bar m^2]^2}\,,\\
\phantom{-}\beta_{\lambda_{4,4}}&=(1-4\eta)\bar{\lambda}_{4,4}+\frac{24\pi}{15}\bar\lambda_{4,4}\bar\lambda_1\frac{\eta+5}{[1+\bar m^2]^2}+\frac{8\pi}{15}\bar\lambda_2\bar\lambda_3\frac{\eta+5}{[1+\bar m^2]^3}-\frac{24\pi}{15}\bar\lambda_1^2\bar\lambda_3\frac{\eta+5}{[1+\bar m^2]^4}
\end{cases}
\label{system2phi8}
\eea
with the anomalous dimension given by equation \eqref{flowZ3}. One more time, the system can be solved numerically, and the fixed points as well as their essential properties are summarized in the Table \eqref{table2}.

\begin{table}
\begin{center}
\begin{tabular}{|l|l|l|l|l|l|l|l|l|l|l|}
\hline FP & $FP_1$&$FP_2$&$FP_3$ &$FP_4$&$FP_5$&$FP_6$&$FP_7$ &$FP_8$&$FP_9$&$FP_{10}$\\
\hline  $\bar{m}^2$ &-1.07&-0.91 &-0.84&-0.84  &-1.22  &-0.75&-0.76 &-0.74&-0.59&1.45\\
\hline  $\bar{\lambda}_{1}$ &0.004&0.005&0.02&0.02&0.009&0.006&0.045&0.05&0.14&-0.65\\
\hline  $10^2\bar{\lambda}_2$ &0.1&0.03&0.7&-0.4&0.2&-0.01&2&1&-40&-16 \\
\hline  $10^3\bar{\lambda}_3$ &-0.3&0.04&-3&0.&-0.7&0.02&-10&-10&-0.6&-3\\
\hline  $10^2\bar{\lambda}_{4,1}$ &0.01&-0.04&0.1&0.01&0.001&-230&0.3&0.3&0.09&-0.3\\
\hline  $10^3\bar{\lambda}_{4,2}$&-0.04&-0.01&0.&-1&-0.06&-2000.&-0.9&-1&100&-10\\
\hline  $10^3\bar{\lambda}_{4,3}$&0.&0.&-0.3&-0.9&0.&0.&0.&0.&0.&0.\\
\hline  $10^2\bar{\lambda}_{4,4}$&0.003&690&0.02&0.4&0.006&670&0.05&0.2&0.03&-0.03\\
\hline  $\eta$&-6.8&-6.3&-6.0&-6.0&2.6&0.9&-5.9&-5.6&-0.65&-0.66\\
\hline  $\theta^{(1)}$&307.8&289.8&179.6&180.0&-110.7&24.7&142.6&137.9&121.6&3.2\\
\hline  $\theta^{(2)}$&-245.8&-112.0&-12+25i&-45.9&-50+13i&14.5&19+6i&20+4i&30&3.1\\
\hline  $\theta^{(3)}$&173.0&-67.8&-12-25i&31&-50-13i&12.2&19-6i&20-4i&23.6&-3.0\\
\hline  $\theta^{(4)}$&113.6&31+6.7i&21+10i&26&-39&7.7&11.7&13+1.7i&18.9&2.8\\
\hline  $\theta^{(5)}$&77+19i&31-6.6i&21-10i&-22&-33&-6&1.2+10i&13-1.7i&17.2&2.7\\
\hline  $\theta^{(6)}$&77-19i&-28.5&-15&10&24.8&5.7&1.2-10i&11&9.8&1.0\\
\hline  $\theta^{(7)}$&-67&-19&6&6.1&-18&-5&9.8&-5.7&6.3&0.7\\
\hline  $\theta^{(8)}$&7.2&5.8&2.2&4&-2.3&-1.3&5.5&5.4&5.2&0.3\\
\hline
\end{tabular}
\caption{Summary of the properties of the non-Gaussian fixed points in the $\phi^8$ truncation.}\label{table2}
\end{center}
\end{table}
\noindent
Interestingly, note that the line of fixed points has disappeared, that is not a surprise, because such line of fixed points  is generally a pathology of the crude truncation. Among the fixed points listed in the table, only $FP_{5}$, $FP_{6}$ and $FP_{10}$ have $D>0$. The over fixed points have a big critical exponent  and become harmful pathology of the model. \\

\noindent
$\bullet$ The fixed point $FP_5$ has seven irrelevant directions and one relevant direction in the UV, and seems to be an IR fixed point, whose   irrelevant directions span an IR-critical surface with seven dimensions. \\

\noindent
$\bullet$ The fixed point $FP_6$ has five relevant directions and three irrelevant directions in the UV. The relevant directions span an UV-multicritical surface of dimension five. The existence of a such sub-manifold is in accordance with the asymptotic safety of the theory. \\

\noindent
$\bullet$ The fixed point $FP_{10}$ has seven relevant directions and one irrelevant direction in the UV. It correspond to an UV fixed point whose revelant directions span a seven-dimensional UV-critical surface. One more time, the existence of a such manifold seems to be in accordance with a non-trivial asymptotically safe theory. 

\noindent
At this stage, it is not obvious to make contact with the fixed points obtained in the previous truncation. The standard way to highlight these relation is to consider truncation with higher and higher valence, and seek convergence of the fixed points. But in our case, the difficulty of a such computations is very improved by the non-locality of the interactions, and these conclusions have to be confirmed by more finer analysis.
Let us remark that the study of the critical exponent  \cite{Wetterich:1989xg} could upset our analysis for the choice of  the truncations with valances greater than $6$. For instance,   the fixed point $FP_5$ leads to  a very large critical exponent and this  can help to show that the truncation in this order remains non consistent. For the fixed point $FP_{10}$ the critical exponent is small, and adding another interaction of valence more than  8 becomes unnecessary.
Unlike, the fixed point  $FP_{10}$ exhibits several relevant directions and  these do not appear in the previous section by using just the truncation with interaction of valence $6$.

\section{Discussion and conclusion}\label{sec6}
In this paper the renomalization group analysis is applied for just renormalizable $T^6_5$ TGFT model in the deep UV limit. Using the simplest approximation consisting in a truncation around the marginal interactions with respect to the perturbative power counting i.e. around the Gaussian fixed point, we have derived the flow equations for each couplings. Because we have focused our attention on the UV sector, the leading contributions to the flow equations provide to the melonic sector, a consideration which considerably simplify the computation of the flow equations. In a second time, using of the appropriate notion of canonical dimension in the UV, we have translated our flow equation in an autonomous system of differential equations, whose we have computed numerically the fixed points, as well as the behavior of the flow's trajectory around each of them. \\

We have find two type of fixed points. The IR fixed points, whose relevant directions in the IR span an  IR-critical surface, a picture in favor of phase-transitions. This is supported, for two of these IR fixed points, by the negative value of their mass-parameter, and the fact that in their vicinity, the effective action turn to be a Ginsburg-Landau like equation for a $\phi^4$ scalar complex theory, advocating a condensed phase transition interpretation. In opposition, the second type of fixed points are UV critical, and their relevant directions in the UV span critical surface with dimensions higher or equal to two, a picture in accordance with a well-defined and non-trivial behavior in the UV for asymptotically safe theories. In all the case, we observe that anomalous dimensions enhanced or weaken the UV-power counting for relevant operators with respect to the perturbative power-counting, a phenomena which seems to indicate a break-down of our crude truncation in these domains of the phase-space. Moreover, the presence of pathological effects as a line of fixed point seems to confirm  these suspicions, as well as its disappearance in a higher-truncation, while our conclusions about asymptotic safety and IR fixed points remain true. The connection between the new fixed point and these ones obtained in the first truncation remains however unclear at this stage without more control over the approximation procedure.\\


\noindent
Finally, note that in the complementary IR regime, the flow equations receive non-melonic contributions. This is due to the fact that, for a very small cut-off, the sums take values $0$ or $1$. As pointed in \cite{Benedetti:2015yaa}, the appropriate rescalling is provided by the standard power counting, and the flow equation turn as well to be an autonomous system, which can be solved numerically. However, we have to keep in mind that our model is defined on a compact manifold ($U(1)^6$ in the referenced paper, $U(1)^5$ in our case).  Then, no phase transition can be occurs, and all the non-Gaussian fixed point reached to the Gaussian one when the cut-off tend to $0$, except if the radius of the circles tend simultaneously to the infinity.

\section*{Acknowledgments}
We thank all the referees, who have anonymously contributed to the improvement of this work.
D O Samary research at Max-Planck Institute  is supported by the Alexander von Humboldt foundation.

\end{document}